 \documentclass[final,3p,times]{elsarticle}

\usepackage{amssymb}

\usepackage{amsmath,amssymb,graphicx,color,braket,hyphenat,makeidx,xcolor}
\usepackage[ocgcolorlinks,colorlinks=true,linkcolor=blue,citecolor=red,linktocpage=true]{hyperref}

\usepackage{bm}
\usepackage{epsf}
\usepackage{subfigure}
\usepackage{epstopdf}
\usepackage{braket}
\usepackage{enumerate}
\usepackage{soul}

\usepackage[normalem]{ulem}

\newcommand{\stkout}[1]{\ifmmode\text{\sout{\ensuremath{#1}}}\else\sout{#1}\fi}

\DeclareMathOperator\erf{erf}

\newcommand{\ignore}[1]{}

\journal{Physica A}

\begin{document}

\begin{frontmatter}



\title{Micro-reversibility and thermalization with collisional baths}


\author[1]{Jannik Ehrich\corref{cor1}}
\cortext[cor1]{corresponding author}
\ead{jannik.ehrich@uni-oldenburg.de}
\author[2]{Massimiliano Esposito}
\author[3]{Felipe Barra}
\author[4]{Juan M.R. Parrondo}

\address[1]{Institut f\"ur Physik, Carl von Ossietzky Universit\"at, 26111 Oldenburg, Germany}
\address[2]{Complex Systems and Statistical Mechanics, Physics and Materials Science Research Unit, University of Luxembourg, L-1511 Luxembourg, G.D. Luxembourg}
\address[3]{Departamento de F\'isica, Facultad de Ciencias F\'isicas y Matem\'aticas, Universidad de Chile, Santiago, Chile}
\address[4]{Dept. Estructura de la Materia, F\'isica T\'ermica y Electr\'onica and GISC. Universidad Complutense de Madrid. 28040-Madrid, Spain}

\date{\today}

\begin{abstract}
Micro-reversibility plays a central role in thermodynamics and statistical mechanics. It is used to prove that systems in contact with a thermal bath relax to canonical ensembles. However, a problem arises when trying to reproduce this proof for classical and quantum collisional baths, i.e. particles at equilibrium interacting with a localized system via collisions. In particular, micro-reversibility appears to be broken and some models do not thermalize when interacting with Maxwellian particles. We clarify these issues by showing that micro-reversibility needs the invariance of evolution equations under time reversal plus the conservation of phase space volume in classical and semiclassical scenarios. Consequently, all canonical variables must be considered to ensure thermalization. This includes the position of the incident particles which maps their Maxwellian distribution to the effusion distribution. Finally, we show an example of seemingly plausible collision rules that do not conserve phase-space volume, and consequently violate the second law.

\end{abstract}


\end{frontmatter}



\noindent
{\em This work is dedicated to the memory of Christian Van den Broeck.}

\section{Introduction}


 Since the inception of statistical mechanics, microscopic reversibility has played a central role. On the one hand, it may seem to contradict the very possibility of irreversible behavior to occur. This was indeed the basis of Loschmidt's criticism of Boltzmann's explanation of macroscopic irreversibility. On the other hand, it is used to derive crucial results from irreversible thermodynamics such as Onsager's celebrated reciprocal relations \cite{deGroot1984} or, more recently, the fluctuation relations for classical \cite{Kawai2007,Gomez-Marin2008} and quantum systems \cite{Campisi2011} which imply the nondecreasing nature of entropy production. 
This underlines the many subtleties which may arise when invoking microscopic reversibility to understand the emergence of irreversibility.

On many occasions, Chris Van den Broeck tackled these issues. As one of the most brilliant representatives of the Brussels school of thermodynamics, Chris was deeply interested in the foundations of statistical mechanics. He contributed to clarifying the relationship between irreversibility and entropy production \cite{Kawai2007,Gomez-Marin2008,Esposito2010,VandenBroeck2010} and exploited Onsager phenomenological equations and fluctuation theorems to find universal properties of the efficiency of thermal and chemical engines \cite{VandenBroeck2005,Esposito2009}. 
In this paper, after revisiting the basic arguments used to prove relaxation to equilibrium based on microscopic reversibility, we consider specifically the issue of thermalization for a fixed system interacting with a bath of thermal particles, where interesting subtleties arise which Chris for sure would have enjoyed.
His remarkable ability to address profound and fundamental problems using simple models has been a guideline for us, and the way we deal here with the problem of micro-reversibility is undoubtedly inspired by Chris' style of doing science. This paper is our tribute to his life and work, which have left a profound mark on the statistical mechanics community. In particular, to ME and JMRP he has been a great mentor as well as a very close friend. 

The first issue that we address concerns the very formulation of the micro-reversibility condition. The generic statement of micro-reversibility is that the probability to observe a transition  $\Gamma\to \Gamma'$  is equal to the probability of the reverse transition $\bar \Gamma'\to \bar \Gamma$. Here, $\Gamma$ and $\Gamma'$ are arbitrary microscopic states of a physical system and $\bar \Gamma$ the time-reversal state of $\Gamma$. The mathematical expression of this statement reads (we provide a more detailed description later on, in Sec.~\ref{sec:db}):
\begin{equation}\label{mr1}
\rho(\Gamma'|\Gamma)=\rho(\bar \Gamma|\bar \Gamma').
\end{equation}
However, this equality immediately poses a problem if the variable $\Gamma$ is continuous. In that case, $\rho(\Gamma'|\Gamma)$ is a density in $\Gamma'$, whereas $\rho(\bar\Gamma|\bar\Gamma')$ is a density in $\bar\Gamma$. Consequently, when the two conditional probabilities are compared, one has to take into account the transformation of the volume elements $d\bar\Gamma$ and $d\Gamma'$. In classical systems, Liouville's theorem warrants the conservation of volume, implying $d\bar\Gamma=d\Gamma'$, and resolves the problem. But micro-reversibility is also relevant for quantum and semi-classical systems with states parametrized by continuous variables $\Gamma$, such as Wigner distributions in phase space or wave packets centered around a given position and velocity \cite{Littlejohn1986,Zachos2005}. In the first case, for instance, it has been shown that there is no equivalent  Liouville-like theorem, that is, unitary evolution does not necessarily conserve the phase-space volume \cite{Sala1993}. Therefore, it is necessary to explicitly check the micro-reversibility condition, Eq.~\eqref{mr1}, in those quantum and semi-classical scenarios.

The second issue concerns the role of micro-reversibility in  the relaxation of a system towards equilibrium. For a classical system in contact with a thermal bath, the micro-reversibility condition applies to micro-states $\Gamma=(x,y)$ of the global system, consisting of the system itself ($x$) and the bath ($y$).
The bath variables can be eliminated by multiplying Eq.~\eqref{mr1} by the equilibrium distribution $\rho_{\rm eq}(y)$ and integrating over $y$ and $y'$. This procedure, which we describe in detail in Sec.~\ref{sec:therm}, is especially relevant for the so-called \emph{collisional baths}, where the system is a fixed scatterer that interacts with particles coming from a thermal reservoir. Nowadays this scenario is realizable in the laboratory. A celebrated example is the interaction between light and atoms in cavity quantum electrodynamics~\cite{Raimond2001}. Moreover, collisional baths are a particular case of a generic type of coupling between a system and a reservoir, consisting of repeated interactions between the system and fresh copies of  small systems called {\em units} or {\em ancillas}, drawn form an infinite reservoir \cite{Scarani2002,Karevski2009,Barra2015,Horowitz2016,Chiara2018}. The thermodynamics of this
system-bath coupling via repeated interactions has been exhaustively studied in Ref.~\cite{Strasberg2017}. However, these previous analyses assume that the interaction is switched on and off by an external agent; then, the system is no longer autonomous and the external agent can perform work. It is worthwhile to go beyond Ref.~\cite{Strasberg2017} and extend the framework  to autonomous collisional baths. The present paper can be considered a first step towards this goal.

Classical collisional baths have been also analyzed in detail as a modification of the Lorentz gas, where the scatterers are fixed disks that can rotate with some angular velocity \cite{Rateitschak2000,Mejia-Monesterio2001,Collet2009,Collet2009b,DeBievre2016}. The collisions between the gas particles and the disks are inelastic because of an exchange of energy between the incident particle and the rotation of the scatterer. This type of models has been used to study different non-equilibrium phenomena like heat transport \cite{Mejia-Monesterio2001,Collet2009,Collet2009b} and thermalization  \cite{Rateitschak2000,DeBievre2016}.

Collisional baths are also interesting regarding the probability distribution of the incident particles. Consider a one-dimensional ideal gas in equilibrium at temperature $T$. The velocity distribution, that is, the probability density for the velocity of a particle chosen at random, is the Maxwellian distribution: $\rho_{\rm Max}(v)\propto e^{-\beta mv^2/2}$, where $m$ is the mass of the gas molecules and $\beta=1/(kT)$, $k$ being the Boltzmann constant. On the other hand, if we sample the velocities of particles hitting a scatterer or crossing a given point, we find that they are distributed according to the {\em effusion distribution} $\rho_{\rm eff}(v)\propto |v|\,\rho_{\rm Max}$. Consequently, a naive treatment would show that the internal degrees of freedom of the scatterer should thermalize if they exchange energy with the incident molecules whose velocity distribution apparently departs from equilibrium. Unfortunately, some of the proposed models, such as the one in Ref.~\cite{Collet2009b}, thermalize when the velocity distribution of the incident particles is indeed Maxwellian, which, at first sight, only adds to the confusion.

In this paper we aim at clarifying these issues by a careful formulation of the micro-reversibility condition and its relation to Liouville's theorem and conservation of phase-space volume in classical and semiclassical collisional reservoirs. 
To this end the paper is structured as follows:
In the following two sections \ref{sec:db} and \ref{sec:mr}, we review the concept of micro-reversibility and its consequences. Specifically, we reproduce a proof of the thermalization of a system subjected to repeated interactions with an equilibrium reservoir. The relation between micro-reversibility and Liouville's theorem is also discussed. 
Section~\ref{sec:discrete} contains the main results of the paper. We analyze a fixed scatterer with discrete states interacting with a generic collisional bath (a detailed description of a physical realization with a semiclassical bath is given in \ref{app:qcollbaths}). We show that it is necessary to take into account the position of the incident particles for a proper formulation of micro-reversibility. Moreover, including this spatial coordinate explains the relevance of the effusion distribution for thermalization. Indeed, failing to do so breaks micro-reversibility in a way which is exactly corrected by the effusion prefactor. 
In Sec.~\ref{sec:class}, we present a classical model with seemingly plausible collision rules which however do not conserve phase-space volume. We show that the system does not thermalize and violates the second law of thermodynamics. 
Finally, in Sec.~\ref{sec:conc} we draw our conclusions.

\section{Micro-reversibility, detailed balance, and equilibrium ensembles}
\label{sec:db}


Before discussing the subtleties associated to  micro-reversibility, in this section we aim to establish a careful mathematical formulation of micro-reversibility and explore its consequences. 
Here, we define micro-reversibility for an arbitrary Markov process $\Gamma(t)$ although, in the next section and the rest of the paper, it will be  applied to classical micro-states and wave packets colliding with quantum scatterers. In the former case, $\Gamma(t)$ consists of all coordinates and momenta and the evolution of $\Gamma(t)$ is deterministic (but still Markovian) and preserves the phase-space volume, according to Liouville's theorem, whereas in the latter case $\Gamma(t)$ consists of the center and momentum of  incident wave packets plus the internal state of the scatterer (see Sec.~\ref{sec:discrete} and \ref{app:qcollbaths}).

To properly formulate the micro-reversibility condition, we need to define two important concepts. First, the conditional probability $\rho(\Gamma'|\Gamma;\Delta t)$, which is the probability (or the probability density) to observe the system in state $\Gamma'$ at time $t+\Delta t$ if the state of the system was $\Gamma$ at time $t$. Along the paper the time interval $\Delta t$ is arbitrary and fixed. We will assume that this conditional probability neither depends on time $t$ (stationarity) nor on the previous history (Markovianity). Notice also that, if the phase space is continuous, then $\rho(\Gamma'|\Gamma;\Delta t)$ is a density only with respect to the final state $\Gamma'$.

The second concept is the time reversal of state $\Gamma$, which we denote by $\bar\Gamma$. It results from $\Gamma$ by inverting velocities and other variables which are odd under time reversal, implying $\bar{\bar\Gamma} = \Gamma$. We assume that every state $\Gamma$ has a  unique time-reversed state.

With these definitions, the micro-reversibility condition can be formulated as:
\begin{equation}\label{micro-rev}
\rho(\Gamma'|\Gamma;\Delta t)=\rho(\bar\Gamma|\bar\Gamma';\Delta t)\qquad \mbox{for all $\Gamma$ and $\Gamma'$}.
\end{equation}
We stress that the above probabilities can denote the mapping of classical deterministic Hamiltonian dynamics (in which case they are delta functions), transition probabilities resulting from unitary quantum evolution, or transition probabilities between the states of a Markov chain (as is commonly assumed in stochastic thermodynamics).

In the following, we will assume that Eq.~\eqref{micro-rev} holds and explore its consequences; specifically, we show that micro-reversibility implies that the corresponding stationary probability distribution $\rho_{\rm st}(\Gamma)$ is compatible with the common equilibrium ensembles: microcanonical for isolated systems and canonical for systems in contact with thermal baths.

In order to do this, we need the evolution master equation  for the probability $\rho(\Gamma,t)$ of observing the system in state $\Gamma$ at time $t$:
\begin{equation}
\rho(\Gamma,t+\Delta t)=\int d\Gamma' \rho(\Gamma',t)\rho(\Gamma|\Gamma';\Delta t).
\end{equation}
Taking into account that $\rho(\bar\Gamma'|\bar\Gamma;\Delta t)$ is normalized with respect to $\bar\Gamma'$ for all $\bar\Gamma$ and that $d\Gamma'=d\bar\Gamma'$, the master equation can be written in the form
\begin{equation}\label{master}
\rho(\Gamma,t+\Delta t)-\rho(\Gamma,t)=\int d\Gamma'\left[\rho(\Gamma',t)\rho(\Gamma|\Gamma';\Delta t)-\rho(\Gamma,t)\rho(\bar\Gamma'|\bar\Gamma;\Delta t)\right]
\end{equation}
and  the stationary solution $\rho_{\rm st}(\Gamma)$ 
satisfies
\begin{equation}\label{jstat}
\int d\Gamma'\left[\rho_{\rm st}(\Gamma')\rho(\Gamma|\Gamma';\Delta t)-\rho_{\rm st}(\Gamma)\rho(\bar\Gamma'|\bar\Gamma;\Delta t)\right]=0.
\end{equation}
One particular way of fulfilling Eq.~\eqref{jstat} is the so-called {\em detailed balance}  condition, that is:
\begin{equation}\label{db}
\rho_{\rm st}(\Gamma')\rho(\Gamma|\Gamma';\Delta t)-\rho_{\rm st}(\Gamma)\rho(\bar\Gamma'|\bar\Gamma;\Delta t)=0
\end{equation}
for any pair of states $\Gamma$ and $\Gamma'$. In the absence of odd variables under time reversal, $\bar\Gamma=\Gamma$ and detailed balance implies that local probability currents vanish.

Inserting Eq.~\eqref{micro-rev} into the detailed balance condition, Eq.~\eqref{db}, one finds
\begin{equation}\label{current2}
\left[\rho_{\rm st}(\Gamma')-\rho_{\rm st}(\Gamma)\right]\rho(\Gamma|\Gamma';\Delta t)=0.
\end{equation}
Therefore, a stationary solution of Eq.~\eqref{master} satisfying detailed balance is the uniform distribution over all connected states, that is: $\rho_{\rm st}(\Gamma')=\rho_{\rm st}(\Gamma)$ for all $\Gamma'$ and $\Gamma$ such that $\rho(\Gamma|\Gamma';\Delta t)\neq 0$.  For ergodic Hamiltonian systems, the set of connected states is an energy layer and the uniform solution is simply the microcanonical ensemble.

\subsection{Thermalization via repeated interactions}
\label{sec:therm}

Consider now a system in contact with a reservoir. We denote by $\Gamma\equiv (x,y)$ the state of the global system, $x$ being the state of the system and $y$ the state of the reservoir. The probability to observe a transition $x\to x'$ in the system in an interval $\Delta t$ reads
\begin{equation}\label{condx}
\rho(x'|x;\Delta t)=\iint dy\,  dy' \rho(x',y'|x,y;\Delta t)\rho(y,t).
\end{equation}
A common assumption in statistical mechanics is that, in each time interval of duration $\Delta t$, the system interacts with a fresh copy of the reservoir in equilibrium, i.e., $\rho(y,t)=\rho_{\rm eq}(y)$. This assumption is implicit in any approximation yielding Markovian dynamics for the state $x$ of the system. It is exact for collisional baths, as we will see below, and is equivalent to Boltzmann's {\em Stosszahlansatz} or, generically, to neglecting any bidirectional effect between the system and the reservoir, such as re-collisions. Under this assumption, the conditional probability in Eq.~\eqref{condx} is independent of $t$ and reads
\begin{equation}\label{condx2}
\rho(x'|x;\Delta t)=\iint dy\,  dy' \rho(x',y'|x,y;\Delta t)\rho_{\rm eq}(y).
\end{equation}
Using micro-reversibility and expanding with $\rho_{\rm eq}(\bar y')$, we get
\begin{equation}\label{condx3}
\rho(x'|x;\Delta t)=\iint dy\,  dy' \rho(\bar x,\bar y|\bar x',\bar y';\Delta t)\rho_{\rm eq}(\bar y')\,\frac{\rho_{\rm eq}(y)}{\rho_{\rm eq}(\bar y')}.
\end{equation}
For a reservoir in thermal equilibrium at temperature $T=1/(k\beta)$, $k$ being the Boltzmann constant, the ratio between the two distributions verifies 
\begin{equation}\label{db0}
\frac{\rho_{\rm eq}(y)}{\rho_{\rm eq}(\bar y')}=e^{-\beta [e(y)-e(y')]}=e^{-\beta [\epsilon(x')-\epsilon(x)]}\quad \mbox{if\ $\rho(x',y'|x,y;\Delta t)\neq 0$,}
\end{equation}
where $e(y)=e(\bar y)$ is the energy of the state $y$ of the reservoir, $\epsilon(x)=\epsilon (\bar x)$ is the energy of the state $x$ of the system, and we have assumed that the evolution conserves the total energy: $e(y)+\epsilon(x)=e(y')+\epsilon(x')$ for any transition $(x,y)\to (x',y')$ with non-zero probability. Inserting Eq.~\eqref{db0} into Eq.~\eqref{condx3}, one immediately obtains
\begin{equation}\label{db1}
\frac{\rho(x'|x;\Delta t)}{\rho(\bar x|\bar x';\Delta t)}=e^{-\beta [\epsilon(x')-\epsilon(x)]},
\end{equation}
which is also known as the detailed balance condition for systems in contact with thermal baths.  Eq.~\eqref{db1} ensures thermalization, that is, one can show that Eq.~\eqref{db1} implies  Eq.~\eqref{db}, changing $\Gamma$ by $x$, if the stationary distribution is the Gibbs state  $\rho_{\rm st}(x)\propto e^{-\beta\epsilon(x)}$.

\section{Proof of micro-reversibility in classical and quantum systems}
\label{sec:mr}

We now address the interpretation of the micro-reversibility condition in Eq.~\eqref{micro-rev}. As mentioned in the introduction, this condition immediately poses a problem if the variable $\Gamma$ is continuous. In that case, $\rho(\Gamma'|\Gamma;\Delta t)$ is a density with respect to $\Gamma'$, whereas $\rho(\bar\Gamma|\bar\Gamma';\Delta t)$ is a density with respect to $\bar\Gamma$. Consequently, when the two conditional probabilities are compared, one has to take into account the transformation of the volume elements $d\Gamma$ and $d\Gamma'$.

In the classical case, Liouville's theorem resolves this issue as it is a central ingredient in the derivation of the micro-reversibility condition. In classical mechanics the conditional probability is a delta function:
\begin{equation}
\rho(\Gamma'|\Gamma;\Delta t) = \delta(\Gamma' - U(\Gamma)),
\end{equation}
where $U(\Gamma)$ is the time evolution operator during the interval $[t,t+\Delta t]$. Since the Hamiltonian evolution in phase space is invariant under time reversal, this operator verifies $\Gamma'={U}(\Gamma)\Rightarrow \bar \Gamma={U}(\bar \Gamma')$. Using the properties of the Dirac delta, we obtain
\begin{align}
\rho(\Gamma'|\Gamma;\Delta t) &= \left|\frac{d \Gamma}{d \Gamma'} \right| \delta( \Gamma - U^{-1}( \Gamma')) \nonumber \\
 &=\left|\frac{d \Gamma}{d \Gamma'} \right| \delta(\bar \Gamma - U( \bar \Gamma')) 
  = \rho(\bar{\Gamma}|\bar{\Gamma}';\Delta t).
\end{align}
In the last step we have used the fact that the Jacobian of the transformation $\Gamma'\to \Gamma$ is $|d\Gamma/d\Gamma'|=1$, by virtue of Liouville's theorem. Notice however that the time-reversal invariance of Hamiltonian evolution alone is not enough to ensure micro-reversibility. Liouville's theorem is crucial in the derivation and we recall that the theorem is valid only if $\Gamma$ is a complete set of canonical coordinates and canonical momenta of the system and the bath. This will be of extreme importance as we will see in the next section. An analogous derivation of micro-reversibility for Hamiltonian dynamics can be found in Ref.~\cite{vanKampen2007}.

In quantum mechanics the proof of the micro-reversibility condition is rather simple at first sight. Let ${\cal U}$ be the unitary evolution operator between $t$ and $t+\Delta t$:
\begin{equation}\label{qprob}
\rho(\Gamma'|\Gamma;\Delta t)=|\braket{\Gamma'|{\cal U}|\Gamma}|^2=|\braket{\Gamma|{\cal U}^\dagger|\Gamma'}|^2.
\end{equation}
The anti-linear time reversal operator $\Theta$ verifies $\Theta^2={\mathbb I}$,  $\Theta^\dagger {\cal U} \Theta={\cal U}^{-1}={\cal U}^\dagger$, and $\ket{\bar \Gamma}=\Theta\ket{\Gamma}$. Therefore:
 \begin{equation}
\rho(\Gamma'|\Gamma;\Delta t)=|\braket{\Gamma|\Theta^\dagger {\cal U} \Theta |\Gamma'}|^2=|\braket{\bar \Gamma| {\cal U} |\bar \Gamma'}|^2 
=\rho(\bar \Gamma|\bar \Gamma';\Delta t). \label{qmr}
\end{equation}
However, this is valid only if $\ket{\Gamma}$ and $\ket{\Gamma'}$ are normalized quantum states and when the labels $\Gamma$ and $\Gamma'$ are discrete. In scattering theory, on the other hand,  $\ket{\Gamma}$ is either a non-normalizable scattering state, where $\Gamma$ denotes the velocity of a plane wave, or it is a normalized wave packet, in which case $\Gamma$ represents the position and the velocity of the packet.  Either way, $\Gamma$ is continuous and micro-reversibility requires the Jacobian of the transformation $\Gamma\to \Gamma'$ to be equal to one. This requirement is generically related to the validity of Liouville's theorem for quantum states parametrized by continuous variables, such as Wigner distributions in phase space or wave packets \cite{Littlejohn1986,Zachos2005}. Since Liouville's theorem is not valid in general for this type of representation \cite{Sala1993}, it is necessary to check the micro-reversibility condition in each case. We partly address this task in the next section.

%
%

\section{Scatterers with discrete internal states: Effusion vs. Maxwellian velocity distributions}
\label{sec:discrete}

\begin{figure}[h]
\[
\raisebox{27pt}{\includegraphics[width=0.4 \linewidth]{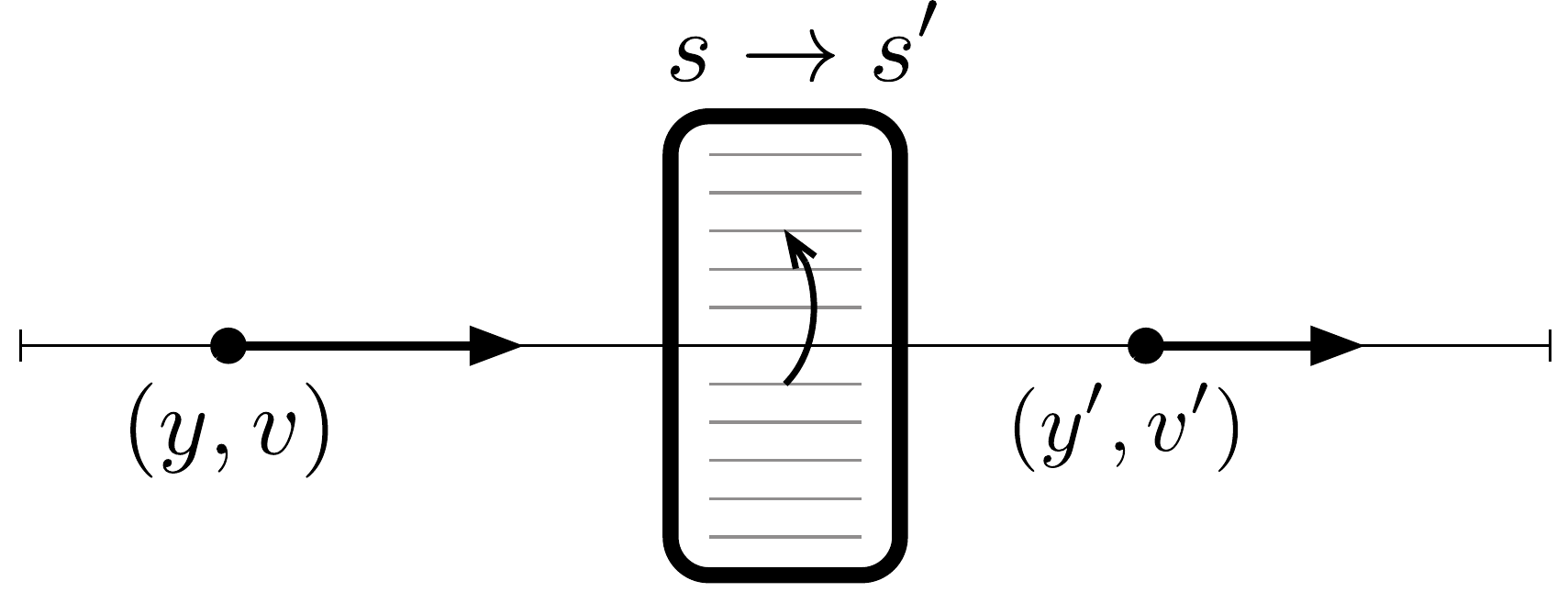}}\qquad\qquad
\includegraphics[width=0.4 \linewidth]{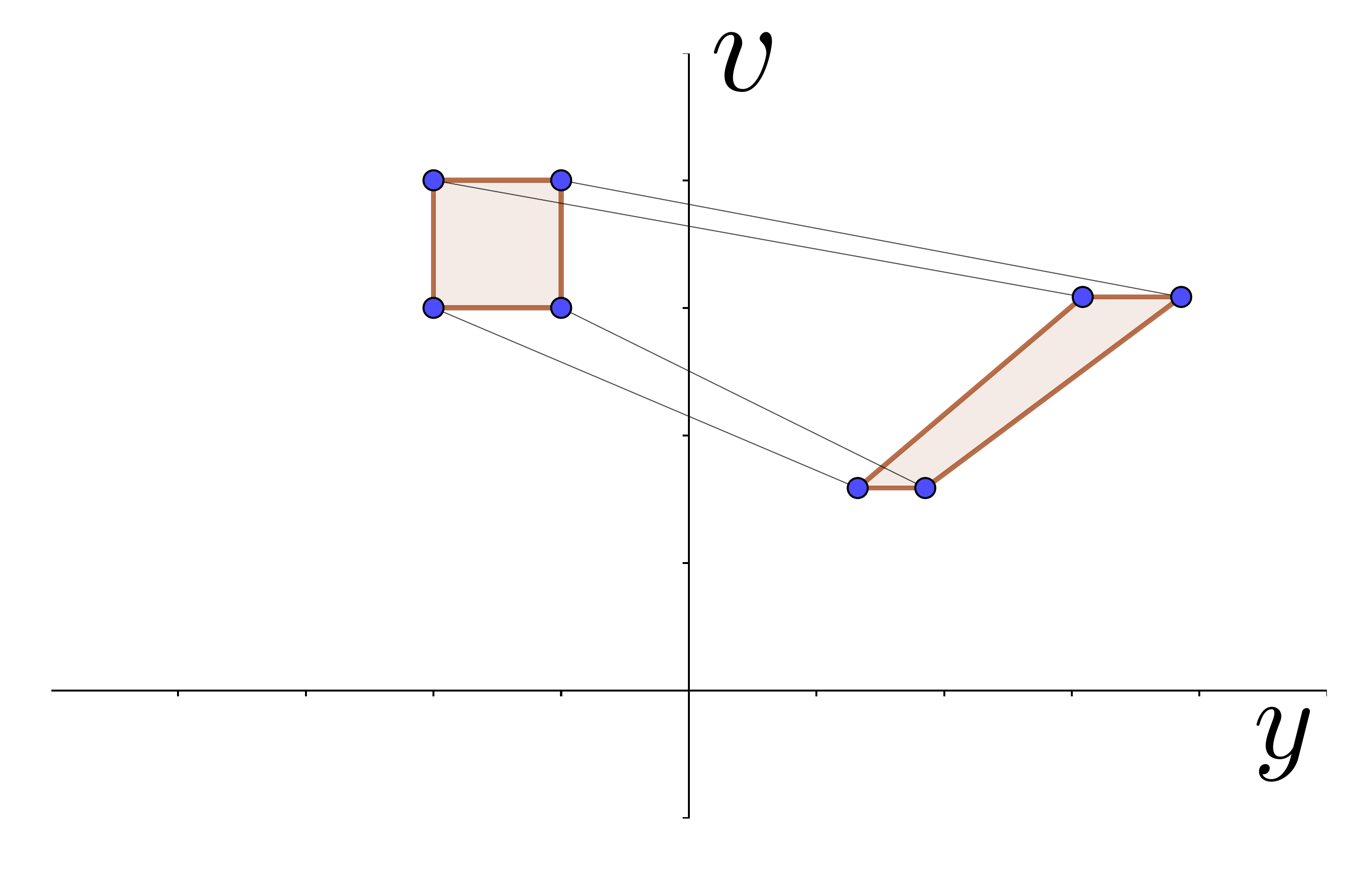}
\]
\caption{{\em Left:} Schematic representation of a collisional bath. A bath particle starts at position $y$ with velocity $v$ and $(y',v')$ are its position and velocity after a time $\Delta t$ during which a collision could occur or not. If a collision occurs, the system changes its internal state from $s$ to $s'$. {\em Right:} Transformation of velocity and position of the bath particles in a collision: $(y,v)\to (y',v')$ as given by Eq.~\eqref{transform} with $2m(\epsilon_{s'}-\epsilon_s)=6.48$ and $\Delta t=1.5$. The plot shows that the transformation conserves the volume of sets in the phase space spanned by $(y,v)$.}
\label{fig:scat1}
\end{figure}

As explained in the introduction, a naive application of the results of the two previous sections to collisional reservoirs leads to unexpected results. In those situations, particles of mass $m$ extracted from an ideal gas in equilibrium at temperature $T$ interact with a fixed scatterer with internal degrees of freedom, exchanging energy, as depicted in Fig.~\ref{fig:scat1} ({\em left}).
Recall that the  distribution $\rho(v)$ of the velocity of the particles hitting the scatterer is the effusion distribution, which is proportional to $|v|e^{-\beta mv^2/2}$ and  does not fulfill the condition given by Eq.~\eqref{db0}. This invalidates the whole argument in Sec.~\ref{sec:therm} such that, superficially, the detailed balance condition in Eq.~\eqref{db1} should no longer hold. Therefore, the system apparently does not thermalize.

The reason for this anomaly is that we have omitted the position of the reservoir particles. As we will see, taking into account the spatial coordinate $y$  of the incident particles is necessary for a proper formulation of micro-reversibility due to Liouville's theorem~\cite{Tolman1938}. Luckily, the inclusion of the spatial coordinate also provides an explanation for why the effusion distribution is the one ensuring  the thermalization of the system.

To discuss this issue in detail, consider a one-dimensional scenario where a fixed scatterer is located at the origin, $y=0$, and has a discrete set of internal states $s=1,2,\dots$ with energy $\epsilon_s$. 
This tacitly implies that the scatterer is quantum and the incident particles must be quantum wave packets. However, the quantum nature of the setup is not relevant in the discussion, which only relies on the transition probabilities $\rho(\Gamma'|\Gamma)$, where $\Gamma$ now comprises the internal state of the scatterer and the parameters defining the incident and outgoing  wave packets.
 For simplicity, we will assume that the scatterer is symmetric, that is, its behavior is identical for particles hitting from the
left  with positive velocity $v>0$ and from the right with negative velocity $-v$. 
Then, we can focus on particles coming from the left, $y<0$, with positive velocity $v$, and exchanging energy with the scatterer. We will further assume 
that there are no reflected particles, that is, the outgoing velocity $v'$ has the same sign as $v$ (see \ref{app:qcollbaths} for a more general case). This setting is depicted in 
Fig.~\ref{fig:scat1} ({\em left}). We denote by $p(s'|s,v)$ the probability of a transition $(s,v)\to s'$. If  the collision is invariant under time reversal, we have the following symmetry in the transition probabilities:
\begin{equation}\label{eqn_microRev_semiclass}
p(s'|v,s)=p(s\,|-v_{s'},s')=p(s\,|v_{s'},s')\qquad \mbox{with $v_{s'}\equiv +\sqrt{v^2-2m(\epsilon_{s'} -\epsilon_s)}$.}
\end{equation}
In \ref{app:qcollbaths} we present a detailed description of a semi-classical collisional reservoir where this symmetry is explicitly derived  using the invariance of the scattering matrix under time reversal. 
Notice that $p(s'|v,s)$ is a probability and not a density. The corresponding probability density for the whole transition $(v,s)\to (v',s')$ reads, for $v,v'>0$:
\begin{equation}
\rho(v',s'|v,s)=\rho(-v',s'|-v,s)=p(s'|v,s)\,\delta\left( v'-\sqrt{v^2-2m(\epsilon_{s'} -\epsilon_s)}\right).
\end{equation}
Again, this transition probability density does not obey the micro-reversibility condition:
\begin{eqnarray}
\rho(-v,s\,|-v',s')&=&p(s\,|v',s')\,\delta\left( v-\sqrt{v'^2-2m(\epsilon_{s} -\epsilon_{s'})}\right)\nonumber \\
&=&
p(s'|s,v)\,\frac{\delta\left( v'-\sqrt{v^2-2m(\epsilon_{s'} -\epsilon_{s})}\right)}{|\partial v'/\partial v|} \nonumber \\ &=&
\left|\frac{v'}{v}\right|\rho(v',s'|v,s).
\end{eqnarray}

That means that a naive analysis of the thermalization process due to collisional baths reveals that neither micro-reversibility is obeyed, nor is the system interacting sequentially with an equilibrium reservoir.

As already anticipated, we can restore micro-reversibility by including the position  of the reservoir particles. In order to do that, we have to calculate the conditional probability  $\rho(y',v',s'| y,v,s;\Delta t)$, where we have included again the explicit dependence on the time interval $\Delta t$. To calculate this conditional probability, consider a reservoir particle initially located at position $y<0$ and with velocity $v>0$, as in Fig.~\ref{fig:scat1} ({\em left}). In an interval $\Delta t$ a collision occurs only if $y+v\Delta t>0$. It then happens at time $t_c=-y/v$, and the final position is $y'=v'(\Delta t- t_c)=v'(\Delta t+y/v)$. In this collision, the transformation in the full phase space reads
\begin{equation}
\begin{cases}
\displaystyle
y'& \displaystyle=\sqrt{v^2-2m(\epsilon_{s'} -\epsilon_s)} \left(\Delta t+\frac{y}{v}\right) \\[0.2cm]
v' & = \sqrt{v^2-2m(\epsilon_{s'} -\epsilon_s)}
\end{cases}\label{transform}
\end{equation}
and its Jacobian is
\begin{equation}\label{jac1}
\left|\frac{\partial (y',v')}{\partial (y,v)}\right|=\left|\begin{array}{cc}v'/v & -yv'/v^2+y/v'+v\Delta t/v'\\0 & v/v'\end{array}\right|=1
\end{equation}
for all $s,s'$. The conservation of phase space volume in this transformation is sketched in Fig.~\ref{fig:scat1} ({\em right}). It is the same conservation that one can observe  in a stream of classical particles 
that change their velocity due to, for instance, a step potential of height $\Delta V$. The potential accelerates or slows down particles exactly as in our discussion, i.e.,  $v'=\sqrt{v^2-\Delta V}$ and, in this case, Eq.~\eqref{jac1} is indeed the expression of Liouville's theorem.

Taking into account the condition for a collision, $y+v\Delta t>0$, the conditional probability is given by
\begin{equation}
\rho(y',v',s'| y,v,s;\Delta t) 
=\begin{cases}
\rho(v',s'| v,s)\,\delta\left(y'-v'\Delta t-{yv'}/{v}\right) &  \mbox{ if  $ 0<-{y}<v\Delta t$}
 \\[0.2cm]
\delta(v'-v)\delta_{s's}\delta(y'-y-v\Delta t) &  \mbox{{otherwise}}
\end{cases}
\end{equation}
and it is straightforward to check that it verifies the micro-reversibility condition, since the Jacobian of the transformation $(y,v)\to (y',v')$ is one. Notice however that, if particles are confined in a container of length $2L$ and the scatterer is in the middle, this expression is valid  only for $|v|\Delta t<L$ . 

Having discussed the subtleties that arise from applying the micro-reversibility condition to collisional baths, we proceed with a \emph{bottom-up} derivation of thermalization of systems with internal degrees of freedom in collisional baths:
The probability of a transition in the system due to an energy exchange with a bath particle during a collision reads
\begin{equation}
\rho(s'|s;\Delta t)=\iint dydy'\iint dvdv' \rho_{\rm eq}(y,v)\rho(y',v',s' | y,v,s). \label{trans-coll-bath}
\end{equation}
Since the bath particles are in equilibrium, their positions and velocities are distributed according to $\rho_{\rm eq}(y,v)=\rho_{\rm Max}(v)/(2L)$ with
\begin{equation}
\rho_{\rm Max}(v)\equiv \sqrt{\frac{\beta m}{2\pi}} e^{-\beta mv^2/2}
\end{equation}
being the Maxwellian distribution in one dimension. The discussion in Sec.~\ref{sec:therm} ensures that $\rho(s'|s;\Delta t)$ satisfies the detailed balance condition in Eq.~\eqref{db1} and the system thermalizes to the Gibbs state $p_{s}^{\rm st}\propto e^{-\beta\epsilon(s)}$.

If one wants to eliminate the positions  $y$ and $y'$ of the incident particles from Eq.~\eqref{trans-coll-bath}, it is more convenient to calculate the transition probability conditioned on the occurrence of a collision, that is $\rho(s'|s;\Delta t)/\pi_c$, where $\pi_c$ is the probability for a collision:
\begin{align}
\pi_c &= 2 \int_{0}^L dy  \int_{-\infty}^{-y/\Delta t}dv\, \frac{\rho_{\rm Max}(v)}{2L}\nonumber \\
 &= \frac{1}{2}\left[ 1- \erf\left( \sqrt{\frac{\beta m}{2}} \frac{L}{\Delta t} \right) \right] + \frac{1-e^{-\frac{\beta m  L^2}{2 \Delta t^2}}}{\sqrt{2 \pi \beta m}}\, \frac{\Delta t}{L} \nonumber \\
 &\approx \frac{\Delta t}{\sqrt{2 \pi \beta m} L} \; \text{as $L \rightarrow \infty$}.
\end{align}
Using Eq.~\eqref{trans-coll-bath}, performing the integral over $y'$, and changing the order of the remaining integrals, the probability of the transition $s\to s'$ in a collision reads
\begin{align}
 p_c(s\to s') =\lim_{L\to \infty} \frac{\rho(s'|s;\Delta t)}{\pi_c} 
&= \frac{\sqrt{2 \pi \beta m}}{\Delta t}\int_0^\infty dv \int_{-\infty}^\infty dv'\int_{-v \Delta t}^0 dy\, \rho_{\rm Max}(v)\rho(v',s'| v,s)\nonumber \\
&=\int_{-\infty}^\infty dv \int_{-\infty}^\infty dv' \sqrt{\frac{\pi \beta m}{2}} |v|\,\rho_{\rm Max}(v) \rho(v',s' |v,s)\nonumber \\
&=\int_{-\infty}^\infty dv \int_{-\infty}^\infty dv' \rho_{\rm eff}(v) \rho(v',s' |v,s),
\end{align}
where
\begin{equation}\label{effd}
\rho_{\rm eff}(v)\equiv\sqrt{\frac{\pi \beta m}{2}} |v|\,\rho_{\rm Max}(v)={\frac{ \beta m}{2}}|v| e^{-\beta mv^2/2}
\end{equation}
is the effusion velocity distribution, that is, the velocity distribution of particles leaving a container through a hole in one of its walls. The effusion distribution is also the probability density of the velocity of particles that cross a point in an equilibrium one-dimensional gas and it is in fact the velocity distribution of the particles hitting the scatterer.

Thus when everything is done right and a complete set of canonical coordinates is chosen, the system thermalizes, as it should be. This is not obvious at first glance, since the system seems to be interacting with a velocity distribution different from the equilibrium Maxwellian distribution. Our analysis shows how this distribution appears: first, one has to consider the position $y$ of the bath particles to properly formulate micro-reversibility; second, once the position $y$ is considered, it is necessary to average over the two variables $(y,v)$ of the bath with respect to the equilibrium state  $\rho_{\rm eq}(y,v)=\rho_{\rm Max}(v)/(2L)$.  The result of this procedure is equivalent to considering incident particles with velocities distributed according to the effusion distribution $\rho_{\rm eff}(v)$. This effect is also important when simulating the thermalization of a particle colliding with a wall acting as a reservoir~\cite{Tehver1998}.

\section{Breaking micro-reversibility: an explicit example}
\label{sec:class}

We now turn to a purely classical example, which is a particular case of the disk scatterers introduced by Mej\'ia-Monasterio, Eckmann, and co-authors in a series of papers \cite{Rateitschak2000,Mejia-Monesterio2001,Collet2009,Collet2009b,DeBievre2016}.  In this particular case the disk scatterer interacts with one-dimensional particles \cite{Collet2009b}. The resulting collision rules comply with the common mechanical requirements -- conservation of energy and angular momentum --, but they are not compatible with conservation of phase-space volume. This has important thermodynamical consequences. For instance, one can build machines beating the second law of thermodynamics. Of course, this indicates that the model is not physically realizable although it looks plausible at first sight. This example illustrates the subtleties that one faces when designing models for collisional reservoirs.

\begin{figure}[h]
\[
\includegraphics[width=0.35 \linewidth]{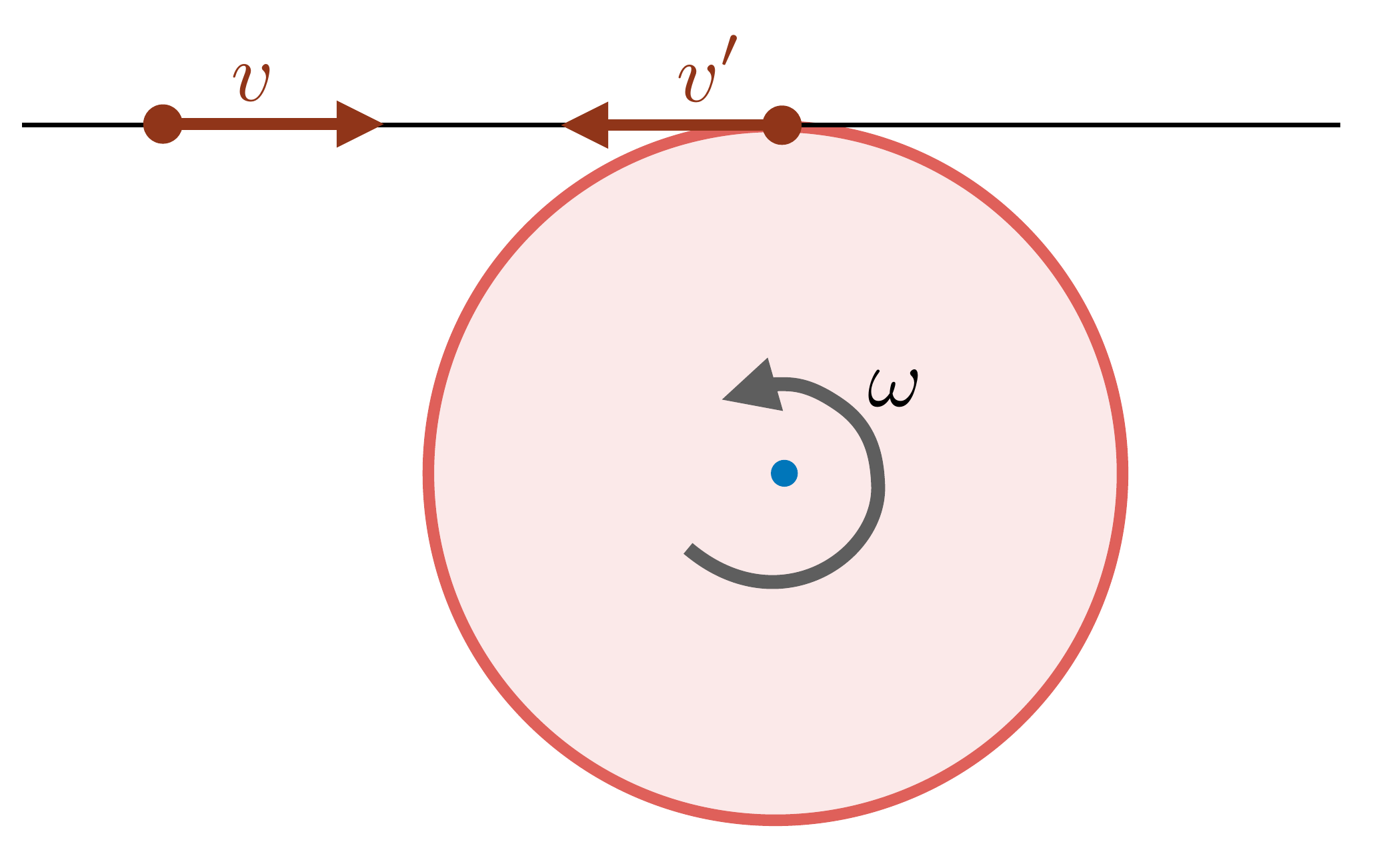}\qquad
\includegraphics[width=0.4 \linewidth]{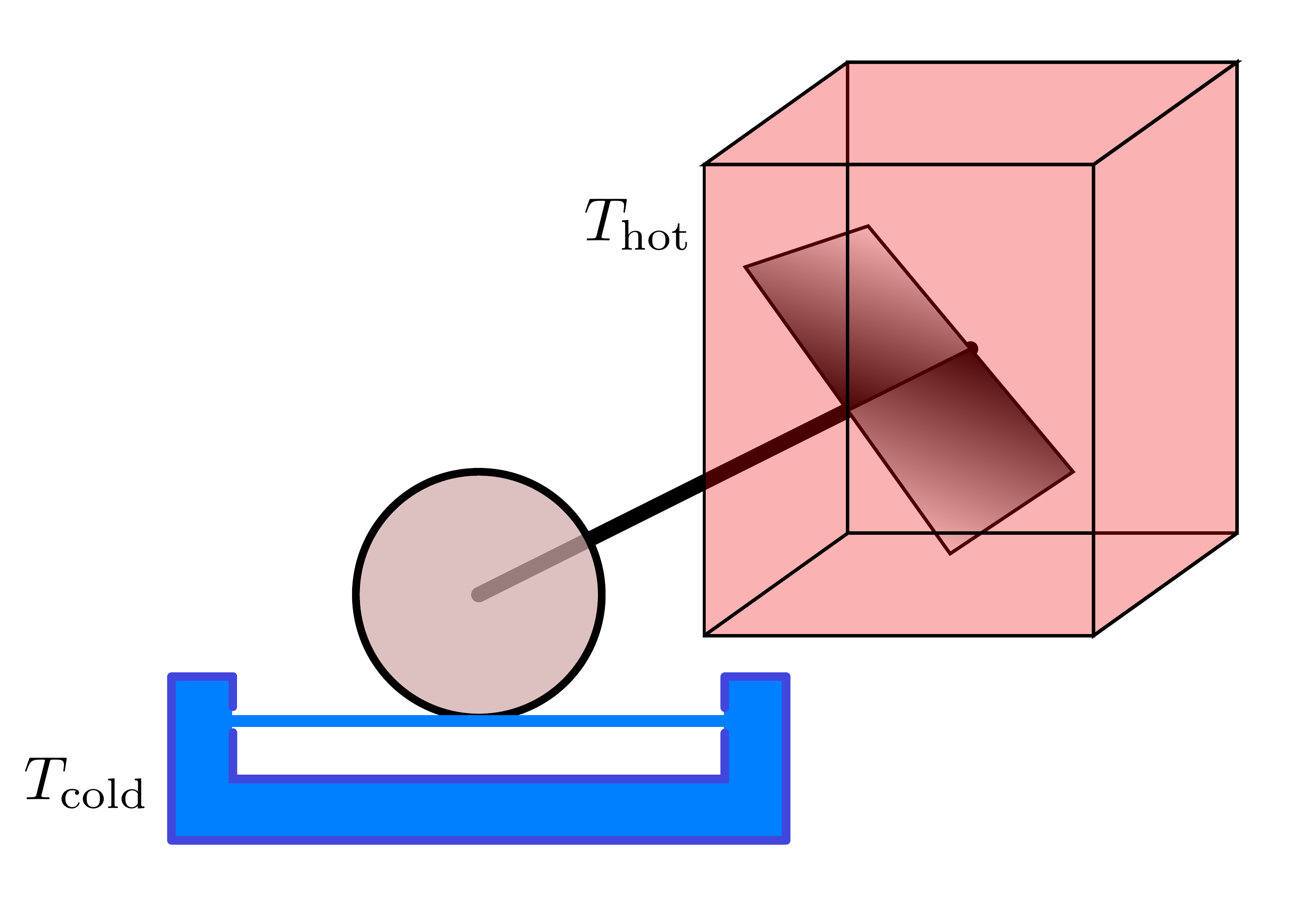}
\]
\caption{{\em Left:} Tangential collision of a particle and  a rotating disk.  If the tangential friction is infinitely strong, there is an exchange of energy between the linear velocity of the incident particle and the angular velocity of the disk given by the transformation \eqref{eqn_disk_coll_vt}. {\em Right:} A ``free'' heat pump consisting of a disk in contact with a one-dimensional cold reservoir and vanes in a hot reservoir. For $\mu=m/I=1$, particles from the cold reservoir exchange velocity with the disk and drive its angular velocity distribution towards the effusion distribution, whose average energy is $kT_{\rm cold}$, whereas the vanes drive the angular velocity to a Maxwellian distribution, with average energy $kT_{\rm hot}/2$. If the temperature of the hot reservoir is such that $T_{\rm cold}<T_{\rm hot}<2T_{\rm cold}$,  heat will be pumped from the cold to the hot bath without the need of external work, in contradiction with the second law of thermodynamics.} 
\label{fig:disk}
\end{figure}

Consider  a homogeneous disk submersed in a one-dimensional ideal gas in equilibrium at temperature $T=1/(k\,\beta)$. The disk has unit radius, a moment of inertia $I$, and rotates with angular velocity $\omega$ about its axis, which is fixed at $(0,0)$. The gas particles have mass $m$ and hit the disk tangentially with velocity $v>0$~[see Fig.~\ref{fig:disk}\,{\em (left)}].

We assume an infinite friction coefficient that allows for the instantaneous exchange of energy between the angular velocity of the disk and the tangential component of the velocity of the incident particle\footnote{Without friction there would be no forces tangential to the disk's surface and thus no change in the angular velocity; this friction, however, does not produce any dissipation and therefore the resulting collisions conserve the total kinetic energy}. In Refs.~\cite{Mejia-Monesterio2001,Collet2009}, these are called ``perfectly rough collisions''. Conservation of energy and angular momentum read
\begin{equation}
\begin{split}
\frac{m}{2} v^2  + \frac{I}{2} \omega^2 &= \frac{m}{2} v'^2 + \frac{I}{2} \omega'^2\\
m v + I \omega &= m v' + I \omega'
\end{split}
\end{equation}
and yield the following collision transformation:
\begin{equation}
\label{eqn_disk_coll_vt}
\begin{split}
v' &= -\frac{1-\mu}{1+\mu} v + \frac{2}{1+\mu} \omega \\
\omega' &=  \frac{2\mu}{1+\mu}v+\frac{1-\mu}{1+\mu} \omega , \end{split}\end{equation}
where $\mu\equiv m/I$. One can easily check that the Jacobian of this transformation is equal to one:
\begin{equation}
J_{v,\omega}=\left| -\frac{1-\mu}{1+\mu}\,\frac{1-\mu}{1+\mu}-\frac{2}{1+\mu} \,\frac{2\mu}{1+\mu}\right|=\left| -\frac{1+\mu^2+2\mu}{(1+\mu)^2}\right|=1 .
\end{equation}
However, one must consider the transformation in the whole phase space $(y,v,\theta,\omega)$, comprising the position $y$ and the velocity $v$ of the incident particle, as well as the rotation angle $\theta$ of the disk and its angular velocity $\omega$:
\begin{equation}\label{coll1d}
\begin{split}
y'&= v'(\Delta t-t_c)\\
v' &= -\frac{1-\mu}{1+\mu} v + \frac{2}{1+\mu} \omega \\
\theta'&=\theta+\omega t_c+\omega'(\Delta t-t_c)\\
\omega' &=  \frac{2\mu}{1+\mu}v+\frac{1-\mu}{1+\mu} \omega\;, \end{split} 
\end{equation}
where $\Delta t$ is the total evolution time and $t_c=-y/v$ is the collision time. The Jacobian of this transformation reads:
\begin{equation}
J=
\left|\frac{\partial (y',v',\theta',\omega')}{\partial (y,v,\theta,\omega)}\right|=\left|\begin{array}{cccc}
\frac{v'}{v} & -\frac{yv'}{v^2}-\frac{1-\mu}{1+\mu}(\Delta t-t_c) & 0 & 0\\[0.2cm]
0 & -\frac{1-\mu}{1+\mu} & 0& \frac{2}{1+\mu} \\[0.2cm]
\frac{\omega'-\omega}{v} & -\frac{y(\omega'-\omega)}{v^2}+\frac{2\mu}{1+\mu}(\Delta t-t_c) &  1 & t_c+\frac{1-\mu}{1+\mu}(\Delta t-t_c)\\[0.2cm]
0 & \frac{2\mu}{1+\mu} & 0 & \frac{1-\mu}{1+\mu}  \end{array}\right| =  \left| \frac{v'}{v}\right|  J_{v,\omega}=\left| \frac{v'}{v} \right|,
\end{equation}
which, in general, is not equal to one.

Therefore, the collision rules in Eq.~\eqref{coll1d} do not conserve phase-space volume and cannot be obtained from a Hamiltonian evolution. This also implies that thermalization is not warranted when the scatterer interacts with particles coming from an equilibrium gas. Indeed, as we have previously discussed,  these particles hit the disk with velocities distributed according to the effusion distribution, Eq.~\eqref{effd}, yielding the following  transition probabilities for the angular velocity:
\begin{align}\label{omomp}
\rho(\omega'|\omega)& = \int_0^\infty dv \,\rho_{\rm eff}(v)\,  \delta\left(\omega' - \frac{1-\mu}{1+\mu} \omega - \frac{2\mu}{1+\mu}v \right)\\
&\propto \rho_{\rm eff}\left(\frac{\omega'-\omega + \mu(\omega'+\omega)}{2 \mu}\right).
\end{align}
The ratio of transition probabilities thus reads
\begin{align}
\frac{\rho(\omega'|\omega)}{\rho(-\omega|-\omega')} = \left|\frac{\omega'-\omega + \mu(\omega'+\omega)}{\omega'-\omega - \mu(\omega'+\omega)}\right| e^{-\beta\, I \left(\omega'^2-\omega^2\right)/2},
\end{align}
that is, the system does not obey detailed balance and consequently the disk does not thermalize. 
This anomaly explains why the authors of Ref.~\cite{Collet2009} find that fixed points in the evolution of angular velocity are obtained by collisions with particles whose velocity distribution is Maxwellian and not, as one would expect, the effusion distribution.

The absence of conservation of phase-space volume can be used to beat the second law. Notice that the average kinetic energy of effusion particles is $kT$, twice the average energy of particles with Maxwellian velocities. Suppose that we place the disk in a one-dimensional gas at temperature $T_{\rm cold}$ and  $\mu=m/I=1$. In this case the incident particle and the disk swap the linear and angular velocity (recall that we take units of length such that $R=1$): $v'=\omega$ and $\omega'=v$. Particles collide with velocities drawn from the effusion distribution, Eq.~\eqref{effd}, and the angular velocity acquires the same effusion distribution, whose average kinetic energy is $kT_{\rm cold}$. Then, we remove the disk and place it in a proper (two-dimensional) thermal reservoir at temperature $T_{\rm hot}\in [T_{\rm cold},2T_{\rm cold}]$. The disk now does thermalize~\cite{Mejia-Monesterio2001,DeBievre2016} and its final average energy is $kT_{\rm hot}/2$. Therefore, it has dissipated an average heat $Q=k(T_{\rm cold}-T_{\rm hot}/2)>0$ to the hot bath. If we repeat the process, in each cycle the disk takes a heat $Q$ from the cold baths and dissipates it to the hot bath without the need of external work, in contradiction to the second law of thermodynamics. One can get a similar heat pump by connecting the disk to vanes immersed in the hot bath, as shown in Fig.~\ref{fig:disk} {\em (right)}.

\section{Conclusions}
\label{sec:conc}

In this paper we have clarified several aspects of the formulation of micro-reversibility and the relationship between micro-reversibility and thermalization. In particular, we have shown the relevance of Liouville's theorem.  We have applied these ideas to one-dimensional semiclassical collisional baths, where reservoir particles exchange energy with a scatterer with discrete internal states. We have found that the collision rules conserve phase-space volume only if the position of the reservoir particles is taken into account and that, in doing so, the effective velocity distribution that thermalizes the scatterer's internal states is the effusion distribution, Eq.~\eqref{effd}. 

Our analysis also reveals that conservation of energy and momentum (or angular momentum) is not enough to ensure the physical feasibility of a given collision rule. In Sec.~\ref{sec:class} we have seen an example of an apparently  sound set of collision rules, Eq.~\eqref{eqn_disk_coll_vt}, which do not conserve phase-space volume and therefore cannot result from Hamiltonian dynamics.
This violation is tantamount to a violation of the second law. Indeed, it  allows one to design a \emph{perpetuum mobile} or ``free'' heat pump, as the one  sketched in Fig.~\ref{fig:disk} {\em (right)}.

We have checked that micro-reversibility is fulfilled by the semi-classical collisional baths introduced in Sec.~\ref{sec:discrete} and \ref{app:qcollbaths} and that, consequently, these baths do induce thermalization. This paves the road to a complete thermodynamic characterization of collisional baths that incorporates the energy exchange between the system and the kinetic energy of the incident particles. In a previous analysis \cite{Strasberg2017}, this energy exchange was interpreted as work performed by an external agent that switches the interaction between the reservoir units and the system on and off. On the other hand, here we have shown that, if the units are particles described by a mixture of wave packets with position and velocity distributed according to the canonical ensemble, this exchange of energy is in fact heat. This distinction is not merely academic, since  work can drive a system out of equilibrium whereas heat induces thermalization. Therefore, each picture yields different qualitative features, particularly for more complicated setups involving several baths or other non-equilibrium sources   
\cite{Barra2015,Chiara2018}. 

%
%
%
%
%
%
%
%
%

\section*{Acknowledgements}

We acknowledge fruitful discussions with Carlos Mej\'ia-Monasterio. JE wishes to thank Andreas Engel for valuable discussions and for enabling a research stay for JE in Madrid.
ME is supported by the European Research Council project NanoThermo (ERC-2015-CoG Agreement No. 681456).
FB acknowledges the financial support of FONDECYT grant 1191441 and of the Millennium Nucleus ``Physics of active matter'' of the Millennium Scientific Initiative of the Ministry of Economy, Development and Tourism (Chile).
JMRP acknowledges financial support from the Spanish Government (Grant Contract, FIS-2017-83706-R). 

%

\appendix

\section{Semi-classical collisional baths}
\label{app:qcollbaths}

\begin{figure}[h]
\[
\includegraphics[width=0.5 \linewidth]{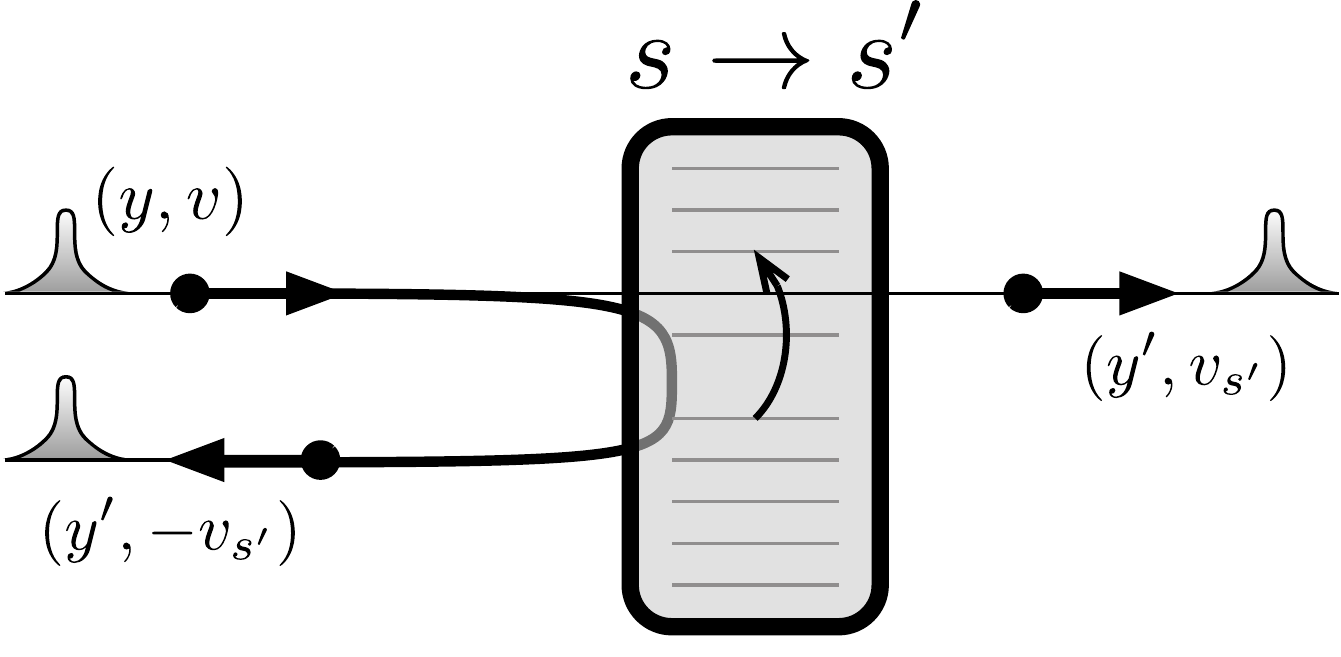}
\]
\caption{Collisional bath consisting of incident wave packets $\ket{v}$ initially located at position $y$. The collision creates a coherent superposition of outgoing wave packets, reflected and transmitted. The coherences are subsequently lost when the packets return to the reservoir.}
\label{fig:packets}
\end{figure}

%
In this appendix, we present an explicit physical realization of the scenario analyzed in section \ref{sec:discrete}: a semi-classical collisional bath corresponding to the following setup [see Fig.~\ref{fig:scat1} ({\em left}) and Fig.~\ref{fig:packets}].
One dimensional quantum particles of mass $m$ extracted from a gas at equilibrium collide with a fixed scatterer located at $y=0$, with  discrete states $s$. The incident particle is described by a wave packet $\ket{v}$ with velocity $v$, and the  state of the scatterer $\ket{s}$ is an eigenstate of the internal Hamiltonian $H_S$. The whole system is described by the Hamiltonian
\begin{equation}
H=\frac{\hat p^2}{2m}+H_S+J(\hat y) H_{S'}=H_0+J(\hat y) H_{S'},
\end{equation}
where $\hat p$ and $\hat y$ are the momentum and position operators on the incident particle, and $H_S$ and $H_{S'}$ are self-adjoint operators in the Hilbert space of the scatterer. The first one is the Hamiltonian of the scatterer with eigenstates $\ket{s}$: $H_S\ket{s}=\epsilon_s\ket{s}$, whereas the second one describes the interaction between the scatterer and the incident particle. Finally, $J(y)$ is a real function whose support is a finite interval $[-l/2,l/2]$, where $l$ is the length of the scattering region. For simplicity, we will consider small scattering regions, such that $l\to 0$. We call this model  semi-classical because the incident particles are described as localized objects at position $y$ with a given velocity $v$. The initial position and velocity of those particles are random and distributed according to some probability density $\rho(y,v)$.

During the collision, the whole system $\ket{v}\otimes \ket{s}$ evolves under a unitary scattering operator  ${\cal S}$ that commutes with the free Hamiltonian $H_0$. In general, the outgoing state ${\cal S} \left[\ket{v}\otimes \ket{s}\right]$ would be the quantum coherent superposition of a number of packets, transmitted and reflected, as sketched in Fig.~\ref{fig:packets}:
\begin{equation}\label{outapp}
{\cal S} \left[\ket{v}\otimes \ket{s}\right]=\sum_{s'}\left[\alpha^{(+)}_{s's}\ket{v_{s'}}\otimes \ket{s'}+\alpha^{(-)}_{s's}\ket{-v_{s'}}\otimes \ket{s'}\right],
\end{equation}
where $\alpha^{(\pm)}_{s's}$ are the amplitudes of transmitted and reflected packets, respectively, and $v_{s'}$ is the positive solution of the conservation of energy condition that follows from $[H_0,{\cal S}]=0$:
\begin{eqnarray}
\epsilon_s+\frac{v^2}{2m}=\epsilon_{s'}+\frac{{v_{s'}}^2}{2m}\Rightarrow v_{s'}\equiv + \sqrt{v^2-2m(\epsilon_{s'}-\epsilon_{s})}.\label{conservenerapp}
\end{eqnarray}

The coherence between those wave packets, however, is lost when the particle returns to the bath. This means that the evolution of the scatterer is ruled by the probability to observe the transition   $(v,s)\to s'$ in a collision. This transition probability is obtained by projecting the outgoing state, Eq.~\eqref{outapp} onto the subspace generated by $\ket{s'}$:
\begin{equation}\label{qprob2app}
p(s'| v,s)=||\bra{s'} {\cal S} \left[\ket{v}\otimes \ket{s}\right]||^2=|\alpha^{(+)}_{s's}|^2+|\alpha^{(-)}_{s's}|^2,
\end{equation}
since $\braket{v\,|-v}=0$ for all $v\neq 0$. It is convenient to split the transition probability into two terms, corresponding to reflected and transmitted wave packets, respectively,
\begin{equation}\label{qprob3app}
p_\pm(s'| v,s)=|\alpha^{(\pm)}_{s's}|^2=|\bra{\pm v_{s'}}\otimes\bra{s'}{\cal S}\ket{v}\otimes \ket{s}|^2.
\end{equation}
The scattering operator ${\cal S}$ is invariant under time reversal, implying
\begin{equation}\label{microrevS1app}
p_+(s'|v,s)=p_-(s|-v_{s'},s'),\qquad p_-(s'|v,s)=p_-(s|v_{s'},s')
\end{equation}
for $v>0$, which correspond, respectively, to transmission and reflection of an incident particle with positive velocity, and
\begin{equation}\label{microrevS2app}
p_-(s'|v,s)=p_+(s|v_{s'},s'),\qquad p_+(s'|v,s)=p_+(s|-v_{s'},s')
\end{equation}
for $v<0$, which correspond to transmission and reflection of an incident particle with negative velocity. Eqs.~\eqref{microrevS1app} and \eqref{microrevS2app}
could in principle be interpreted as an expression of micro-reversibility. However, the probabilities defined in Eq.~\eqref{qprob3app} are not probability densities with respect to the outgoing velocity of the packets. They are in fact probabilities because the states $\ket{v}\otimes \ket{s}$ are normalized.
The complete transition probability then reads
\begin{equation}
\rho(v',s'|v,s) =p_+(s'|v,s)\delta\left( v'-\sqrt{v^2-2m(\epsilon_{s'} -\epsilon_s)}\right)+p_-(s'|v,s)\delta\left( v'+\sqrt{v^2-2m(\epsilon_{s'} -\epsilon_s)}\right),
\end{equation}
which does not verify the micro-reversibility condition. Indeed, using Eq.~\eqref{microrevS1app} and the properties of the Dirac delta, we obtain, for $v>0$:
\begin{eqnarray}
\rho(-v,s|-v',s') &=& p_-(s|-v',s')\delta\left[ -v+\sqrt{v'^2-2m(\epsilon_{s} -\epsilon_{s'})}\right]\nonumber\\
&=& p_-(s|-v',s')\,\frac{\delta\left[ v'-\sqrt{v^2-2m(\epsilon_{s'} -\epsilon_s)}\right]+\delta\left[ v'+\sqrt{v^2-2m(\epsilon_{s'} -\epsilon_s)}\right]}{|\partial v'/\partial v|}\nonumber \\
&=& p_+(s'|v,s)\,\frac{\delta\left[ v'-\sqrt{v^2-2m(\epsilon_{s'} -\epsilon_s)}\right]}{|\partial v'/\partial v|}+
 p_-(s'|v,s)\,\frac{\delta\left[ v'+\sqrt{v^2-2m(\epsilon_{s'} -\epsilon_s)}\right]}{|\partial v'/\partial v|}
\nonumber \\
&=& \left|\frac{v'}{v}\right|\,\rho(v',s'|v,s).
\end{eqnarray}

Notice that micro-reversibility is not fulfilled as, in general, the Jacobian of the transformation $v\to v'$ is not equal to one:  $|dv'/dv|=|v/v'|\neq 1$. On the other hand, if one takes into account the position of the incident particles, as discussed in the main text, the Jacobian of the full transformation $(y,v)\to (y',v')$ is equal to one and micro-reversibility is restored.


\bibliography{referencesThermalization}

\end{document}